\documentclass[useAMS,usenatbib,twocolumn]{mnras}
\usepackage{times,graphicx,amsmath,amsfonts,amssymb}
\usepackage{epsfig}
\usepackage{tabularx}
\usepackage{booktabs}
\usepackage{multirow}

\usepackage{array}
\newcolumntype{L}[1]{>{\raggedright\let\newline\\\arraybackslash\hspace{0pt}}m{#1}}
\newcolumntype{C}[1]{>{\centering\let\newline\\\arraybackslash\hspace{0pt}}m{#1}}
\newcolumntype{R}[1]{>{\raggedleft\let\newline\\\arraybackslash\hspace{0pt}}m{#1}}

\newcommand{\hi}{\mbox{H{\scriptsize I}}}

\def\ltsim{\lower.5ex\hbox{$\; \buildrel < \over \sim \;$}}
\def\gtsim{\lower.5ex\hbox{$\; \buildrel > \over \sim \;$}}
\def\ltsim{\lower.5ex\hbox{$\; \buildrel < \over \sim \;$}}
\def\gtsim{\lower.5ex\hbox{$\; \buildrel > \over \sim \;$}}
\def\gm{\gamma_{_M}}
\def\be{\begin{equation}}
\def\ee{\end{equation}}
\def\ba{\begin{eqnarray}}
\def\ea{\end{eqnarray}} 

\def\kms{\, {\rm km }\, {\rm s}^{-1}}
\def\rtr{r_\mathrm{tr}}
\def\mnras{MNRAS}

\def\dd{\mathrm{d}}
\def\ce{\mathcal{E}}
\def\cem{{\mathcal{E}_\mathrm{m}}}

\def\vp{w}
\def\cL{\mathcal{L}}
\def\vz{u}
\def\vc{V_\mathrm{c}}

\def\ln{{\rm ln}\,}

\def\br{\mathbf{r}}
\def\bv{\mathbf{v}}

\def\bvp{\mathbf{w}}

\def\bz{\mathbf{z}}

\newcommand{\NDF}{\textsc{NGC1052-DF2}}

\title[High  mass of   \NDF ]{Towards a   higher mass for \NDF: an analysis based on full distribution functions }
\author[A. Nusser]{Adi Nusser$^{1,2}$ \thanks{E-mail: adi@physics.technion.ac.il} \\
$^{1}$Department of Physics and the Asher Space Research Institute, Israel Institute of Technology Technion, Haifa 32000, Israel\\
$^{2}$Guangdong Technion-Israel Institute of Technology, Shantou 515063, P.R.China 
}

\begin{document}
\maketitle
\label{firstpage}
\begin{abstract}
It is  demonstrated  that the  kinematics  of  the 10 star clusters in \NDF\  is compatible with  a high dynamical mass  close to those 
implied by  the standard stellar-to-halo-mass ratio (SHMR). The analysis relies on a convenient form for the distribution function (DF) 
of projected phase space data, capturing  non-gaussian features in the spread of true velocities of the mass tracers. 
A key ingredient  is  tidal stripping  by the gravity of the apparently  nearby larger galaxy, NGC 1052. Tidal stripping decreases  the
range of velocities of mass tracers, while  only mildly lowering the total mass inside the trimming radius $\rtr$. 
The analysis is performed assuming halo profiles consistent with simulations of the $\Lambda$CDM model. For the fiducial value 
$\rtr=10$ kpc, we find that the virial mass of the pre-trimmed halo is  $M<1.6\times 10^{10}M_\odot$ at $2\sigma $ ($95\%$)  and 
$M<8.6\times 10^{9}M_\odot$   at $1.64\sigma$ ($90\%$). For the mass within 10 kpc we obtain, $M_\mathrm{10kpc}<3.9\times 10^{9}M_\odot$  
and $<2.9\times 10^{9}M_\odot$  at  $2\sigma$ and $1.64\sigma$, respectively. The $2\sigma$ upper  limit on the virial mass is roughly a 
factor of 3-5  below  the mean SHMR relation.Taking $\rtr=20$ kpc, lowers the $2\sigma$ virial mass limits by  a factor of $\sim 4 $, 
 bringing our results closer to those of Wasserman et al. (2018) without  their SHMR prior.
\end{abstract}

\section{Introduction}
\label{sec:intro}

Assuming it is  at a distance of $\approx 20$ Mpc, the ultra-diffuse  galaxy \NDF\  exhibits several interesting properties. The  galaxy hosts 10 star clusters with luminosities  similar to globular clusters found in brighter galaxies \citep{vanDokkum2018}. Particularly  intriguing is the low velocity dispersion derived from the line-of-sight (l.o.s) of the 
10 star clusters. 
{\b Within the framework of Newtonian gravity,  if  $D\approx 20$ Mpc  then     mass models, based on the \NDF\ kinematics alone,  prefer a  dark matter (DM) halo mass \citep{vanDokkum2018,Wasserman2018} which is substantially low compared   to the virial mass inferred  from  the stellar content of the galaxy.}
{ Indeed, assuming the  standard stellar-to-halo-mass ratio (SHMR) \citep{Behroozi2010,Moster2013,Rodriguez-Puebla2017} , the stellar mass $\sim 2 \times 10^8M_\odot$   of \NDF\ 
implies a virial mass of $\sim 6 \times 10^{10}M_\odot$. }
This places the galaxy at the extreme of galaxy formation models designed in the framework of  the $\Lambda$CDM scenario. 
However, the small number of tracers, missing 3D positions and large l.o.s velocity errors, leave little prospect for 
a robust measurement of the actual mass of the system. It is thus  not surprising  
that authors \citep[e.g.]{vanDokkum2018,Martin2018,Hayashi2018,Trujillo2018,Wasserman2018} quote  90\% upper limits on the virial mass of the \NDF\ halo. Since \NDF\ may well be a satellite galaxy \citep{vanDokkum2018},  It is important to remember  that in the case of satellite
 galaxies, the SHMR gives the virial  mass of the progenitor prior to any tidal stripping by the gravitational tile of the harboring galaxy.
 { Upper limits have also been obtained on the dynamical mass within the projected radius of the most distance tracer, i.e. $R \ltsim 10$ kpc. 
The inferred masses depend on the assumed dark matter halo profiles as well as the assumed distribution of tracers \citep{vanDokkum2018,Hayashi2018,Wasserman2018}. }

In this paper, we continue to ask  whether a high mass close to the SHMR can indeed be ruled out with a high confidence level.
 In contrast to previous studies, we attempt to answer this question using a formalism based on   full distribution functions (DFs)  of the observed 
 projected phase space of tracers.  We also take into account the effect of the  removal of outer parts of the galaxy by the action of external gravitational tides.
We will assume that the \NDF\   is originally associated with a high halo mass, which   had lost (or never accreted)  mass beyond a trimming radius $\rtr $, because of the action of 
the gravitational tidal field of a nearby larger  galaxy. As a fiducial value we consider $ \rtr $, which is consistent with the extent of  both the observed surface brightness  
of \NDF\ and the distribution of the star clusters. We do not require that the mass inside  this radius is decreased and we will assume that it 
remains the same. We explore the virial mass of the original  halo and the mass within $\rtr$, using the projected phase space data of 
projected distances and l.o.s velocities of the 10 star clusters. 
There are several methods  for  inferring  mass from projected phase space data \citep[e.g.][]{2014RvMP...86...47C}. Here we employ certain assumptions that will allow us
to express the distribution function of projected phase space variables in a simple and compact form.  The DF allows a construction of a likelihood function of data, without invoking additional assumptions on moments of the true l.o.s velocities. 

 The distance to \NDF\ is required  to convert  measured angular separations  into 
projected distances of tracers and  to derive the stellar mass 
from the 2D stellar distribution.
There is an ongoing debate regarding the distance to \NDF . The  distance of $D =19\pm 1.7 $ Mpc reported by 
\cite{vanDokkum2018}, has been disputed  by \cite{Trujillo2018} who argue in favor of $D\approx 13 $ Mpc. 
More recently, a new distance of $D=18.7\pm 1.7 $ Mpc is provided by \cite{vanDokkumDist}, consistent with their original estimate.
Deeper observations the color magnitude diagram should yield a  complete resolution to the distance conundrum. 
The nearer distance  leads to a  smaller  stellar mass ($\sim 6\times  10^7 M_\odot$) and, thus, 
 brings \NDF\ closer  to  
 the standard SHMR \citep{Trujillo2018}. 
Therefore, the main interest is in the mass models for  $D= 20$ Mpc and here we opt to present results for this distance only. We only mention that the actual  constraints on the virial halo mass are not  significantly different from $D=13$ Mpc.

 The outline of the remainder  of the paper is as follows. 
 The methodology of constructing the DFs  of the observables is described in \S\ref{sec:obs}. In \S\ref{sec:massmodel},  the mass model of \NDF\ is presented in terms of a stellar component embedded in a DM halo. This section also includes a detailed description of the effect of tidal stripping. The \S\ref{sec:results} gives the upper limits obtained  from our  DF based   likelihood 
function. The paper is concluded with a discussion in \S\ref{sec:discussion}.
  
\section{Observables and distribution functions}
\label{sec:obs}
We assume that the galaxy is in a steady state where the 
 mass distribution is probed  by a sample of tracers with measured 
l.o.s. velocity,  $\vz_0$,   and projected (perpendicular to the l.o.s) distances, $R_0$, with respect to the galaxy center.  The true values of these observables  are denoted simply by $\vz$ and $R$.
We assume that the measurement errors  in $R_0$ are negligible. As in  \cite{Wasserman2018}, we introduce $\vz_\mathrm{sys}$ to allow for 
uncertainties in the determination of  the systemic velocity. Thus,
we write, for  tracer $i$, 
\begin{equation}
\label{eq:Rimodel}
R_{0i}=R_i \, ,
\end{equation}
and
\begin{equation}
\label{eq:uimodel}
\vz_{0i}=\vz_i+\vz_\mathrm{sys}+\delta\vz_i,
\end{equation}
where  $\delta \vz_i$ are the individual measurement errors.  

The likelihood function, $\cL(\vz_0, R_0)$ for observing  $\vz_0$ and $R_0$ is then written as
\begin{equation}
\label{eq:like}
\cL=\int \dd u_\mathrm{sys} P( \vz_\mathrm{sys} )\prod_i \int \dd \vz_i G(\vz_{0i}|{\vz_i+\vz_\mathrm{sys}}) F(\vz_i,R_{0i})\; . 
\end{equation}
We take  the probability distribution function $P( \vz_\mathrm{sys} )$  to be  a Gaussian  with the mean of the observed velocities and a standard deviation of $5\kms$ \citep{Wasserman2018} .
Also $G$ is a Gaussian with  mean $\vz_i+\vz_\mathrm{sys}$ and $\mathrm{rms}$ scatter $\sigma_{ui}$ representing the 
random  error $\delta \vz_i$ in measuring $\vz_{0i}$. The main challenge in computing $\cL$ is
the distribution function (hereafter DF), $F$,  of the underlying quantities $\vz$ and $R$. We describe below our derivation of $F $ from a given  mass model for the galaxy. 
 Constraints on the mass model will then be obtained by maximizing $\cL$ 
  with respect to the parameters of this model.

 The unknown velocity in the projected plane is denoted by $\bvp$ with an amplitude $\vp$. Thus the 3D 
 velocity of a tracer is $\bv\equiv \bvp+\vz\hat{\bz}$, where the l.o.s is arbitrarily chosen as the positive $z$ axis.
Assuming azimuthal  symmetry, 
 $F(\vz,R)$ will be derived from a model for the DF, $f(\bv,\br)$, of the full phase space information of 3D velocities and positions.
This is done,  first by  computing the  DF, $f_u(u,\br)$, of the line-of-sight velocity component and the 3D distance $r$, 
\begin{equation}
\label{eq:f2fu}
f_u(u,\br)=\int \dd^2 \bvp f(\bv,\br)\; .
\end{equation}
 Then $F$  is obtained using
\begin{equation}
\label{eq:fuF}
F(\vz,R)=\int\dd z f_u(\vz,R,z)\; .
\end{equation}

\subsection{The ergodic (isotropic) DF}
\label{sec:ergodic}
We adopt the definitions
\begin{equation}
\ce=\Psi-v^2/2\quad \textrm{and} \quad\Psi=-\Phi\; , 
\end{equation}
where $v^2=\vz^2+\vp^2$ and $\Phi$ is the gravitational potential generated by all components of the system.
We assume now  that  the full phase space DF is of the form $f(\bv,\br)=f(\ce)$. 
This form of the steady state DF implies spherical symmetry as well as  isotropic velocity distribution and is appropriately termed  ``ergodic" by \cite{Binney2008}.
We further impose  $\Psi\rightarrow 0$ as $ r\rightarrow \infty $ and $f(\ce)=0 $ for $\ce\le 0$.
Then, the function $f(\ce)$ obeys \citep[e.g.][]{Eddington16,Binney2008},
\begin{equation} 
\label{eq:feps}
f(\ce)=\frac{1}{\sqrt{8}\pi^2}\frac{\dd}{\dd \ce}\int_0^\ce\frac{\dd \Psi}{\sqrt{\ce-\Psi} }\frac{\dd \nu}{\dd \Psi}\; , 
\end{equation}
where $\nu(r)$ is the distribution of mass  tracers and $\dd \nu /\dd \Psi= (\dd \nu/\dd r)(\dd \Psi/ \dd r)^{-1}$.

The  DF, $f_u$ is obtained from Eq.~\ref{eq:f2fu}, by integrating $f$ over the projected velocity where   $\dd^2 \bvp=2\pi \vp \dd \vp$,
\begin{equation}
f_u(\vz,r) =2\pi\int \vp \dd \vp f(\ce)\; .
\end{equation}
At fixed $r$ and $\vz$ we have $\dd \ce =\vp \dd \vp$ and thus, 
\begin{equation}
f_u(\vz,r) =2\pi\int_0^{\cem}\dd \ce f(\ce)  \; ,
\end{equation}
where 
\begin{equation}
\cem=\Psi(r)-\frac{\vz^2}{2}
\end{equation}
corresponds to $\vp=0$. Together with  Eq.~\ref{eq:feps} we then arrive at \citep[c.f.][]{Dejonghe92}
\begin{equation}
f_u(\vz,r) =\frac{1}{\sqrt{2}\pi}\int_0^{\cem}\frac{\dd \Psi}{\sqrt{\cem-\Psi} }\frac{\dd \nu}{\dd \Psi}\; , 
\end{equation}
offering  a convenient form for deriving $f_u$ from $\Psi$ and $\nu$ under the assumption of ergodicity.
\subsection{DF for circular orbits}
\label{sec:circ}
An ergodic DF corresponds to an  isotropic velocity  dispersion. In complete contrast, is the case of  particles moving  on circular orbits where 
the velocity of a tracer is perpendicular to  $\br$, with a magnitude 
$\vc(r)=\sqrt{GM(r)/r}$.  Let  $\theta$ be  the angle between  $\br$ and the $z$ axis and $\phi$ be the azimuthal angle between the $x$ axis and the projection of $\br$ onto the $xy$ plane. 
Taking  $\eta$ to be the angle between $\bv$ and the azimuthal direction, we find
 \begin{equation}
 \vz=\vc(r) \sin\theta \sin\eta
 \end{equation}
 We write the DF of $u$ and $\br$  as $f_u^\mathrm{circ}(u,\br)=P(u|\br) \nu(r)$ where the probability distribution function (PDF)
 $P(u|\br)$ can be  expressed
 in terms of the  PDF, $P_{_{\!\eta}}$, for $\eta$ as 
\begin{equation}
P(\vz|\br)=P_{_{\!\eta}}\frac{\dd \eta}{\dd \vz}.
\end{equation}
Clearly $P(\vz|\br)=0$ for $|\vz| > \vc \sin\theta$. Otherwise, 
\begin{equation}
P(\vz|\br)=\frac{P_{_{\!\eta}}}{ \sqrt{\vc^2\sin^2\theta-\vz^2}}
\end{equation}
where we have used $\cos\eta=\sqrt{1-\vz^2/(\vc^2\sin^2\theta)}$. A 
 uniform $P_{_{\!\eta}}=const$, yields isotropic velocity dispersion in the tangential direction, i.e. $\sigma^2_{\phi}=\sigma^2_{\theta}$ and corresponds to
  \citep{1987ApJ...313..121M}
\begin{equation}
\label{eq:circiso}
P(\vz|\br)=\frac{1}{\pi}\frac{1}{ \sqrt{\vc^2\sin^2\theta-\vz^2}}\, , 
\end{equation}
which has a  minimum of $P(\vz|\br) $  at $\vz=0$ and diverges  at $|\vz| \rightarrow \vc\sin\theta$. 
{ While spherical
systems have no preferred pole, 
hence $\sigma_\phi=\sigma_\theta$, moving away from spherical symmetry will lead to $\sigma_\phi\ne \sigma_\theta$, so to explore beyond the effects of strict spherical symmetry, we also test
$P_{_{\!\eta}}\ne const$. }
The choice  $P_{_{\!\eta}}\propto |\cos\eta| $ leads to  $\sigma^2_{\phi}=2\sigma^2_{\theta}$ and
\begin{equation}
\label{eq:circ}
P(\vz|\br)=\frac{1}{2\vc \sin\theta}\, , 
\end{equation}
which is uniform for  $|\vz|\le \vc \sin\theta$ and zero otherwise. 

Using $\sin\theta=R/\sqrt{R^2+z^2}$ we write the DF, $f_u^\mathrm{circ}=\nu P(\vz|\br)$, as 
\begin{equation}
f_u^\mathrm{circ}(\vz|\br)=\nu(r)P(\vz|R,z)\; .
\end{equation}
Integrating this function along the l.o.s. as in Eq.~\ref{eq:fuF} gives the corresponding $F(u.R)$.
Note that although $\nu$  is assumed to be  spherically symmetric, the last expression remains valid also for azimuthal symmetry, i.e. 
$\nu=\nu(R,z)$.

\section{Mass model}
\label{sec:massmodel}

We model the galaxy as a two component system of stars and DM, both  assumed spherical. 
The DM component is assumed to follow the   Navarro-Frenk-White (NFW)  density profile  \citep{Navarro1996}, with parameters consistent with  
 the structure of  halos identified in  cosmological simulations \citep[e.g.][]{Gao2008,Ludlow2013,Dutton2014a,Diemer2015}. 
  We define $r_{200}$ as the radius of the sphere  within which the mean halo density is 200 times the critical density
  $\rho_\mathrm{c}=3H_0^2/8\pi G$\footnote{Throughout we take the Hubble constant $H_0=70\kms\; \mathrm{Mpc}^{-1} $. }.   We treat  $r_{200}$ as  the virial radius and the  mass $M=M_{200}$ inside $r_{200}$ as the  virial mass (ignoring  small differences with other definitions of the virial mass).
  In addition to the mass, the second and only other parameter needed to specify the NFW profile is the concentration parameter, $C$. We adopt the fitting formula for the concentration-mass relation, $\bar C(M)$,  found  by  \cite{Dutton2014a} for halos identified in simulations.  

We seek the highest virial halo mass consistent with the \NDF\ kinematics, we will 
consider the lowest concentration which remains compatible with  the scatter around the mean $\bar C(M)$, as  found in simulations.
A concentration of 40\%  below $\bar C(M)$,  falls roughly within a $1\sigma$ scatter  \citep[e.g.][]{Gao2008,Ludlow2013}. Thus,  we will provide results for 
   $C= \bar C(M) $ and also for  lower values $C=0.6\bar C$.  
We could also work with the   Einasto profile which is a more general fit to the halo structure. 
  However, this profile  involves a third parameter, 
  the power index $\alpha_\mathrm{E}$.  Although, for the  relevant mass range,  $\alpha_\mathrm{E}$ has a very weak dependence on mass,  converging to $\alpha_\mathrm{E}=0.167$, the scatter is important and we lack a quantification of the covariance between the 
  concentration and the power index \citep[e.g][]{Dutton2014a}. Therefore, for our purposes here we prefer to work with the 2 parameter 
   NFW fit to describe the DM halo. 
   
 As for the   stellar component, we adopt 
 an Einasto profile with   parameters  $r_{200}=10$ kpc, $C_\mathrm{E}=4$ and $\alpha_\mathrm{E}=1.25$,  which yields a good fit to the two dimensional S\'ersic  profile representing the surface brightness of \NDF \footnote{The author thanks the referee for proposing these parameters} \citep{vanDokkum2018}. The stellar mass is
 is normalized to $2\times 10^8 M_\odot$, in accordance with  the observations.

\begin{figure}
   \vskip 0.2in
 \includegraphics[width=0.48\textwidth]{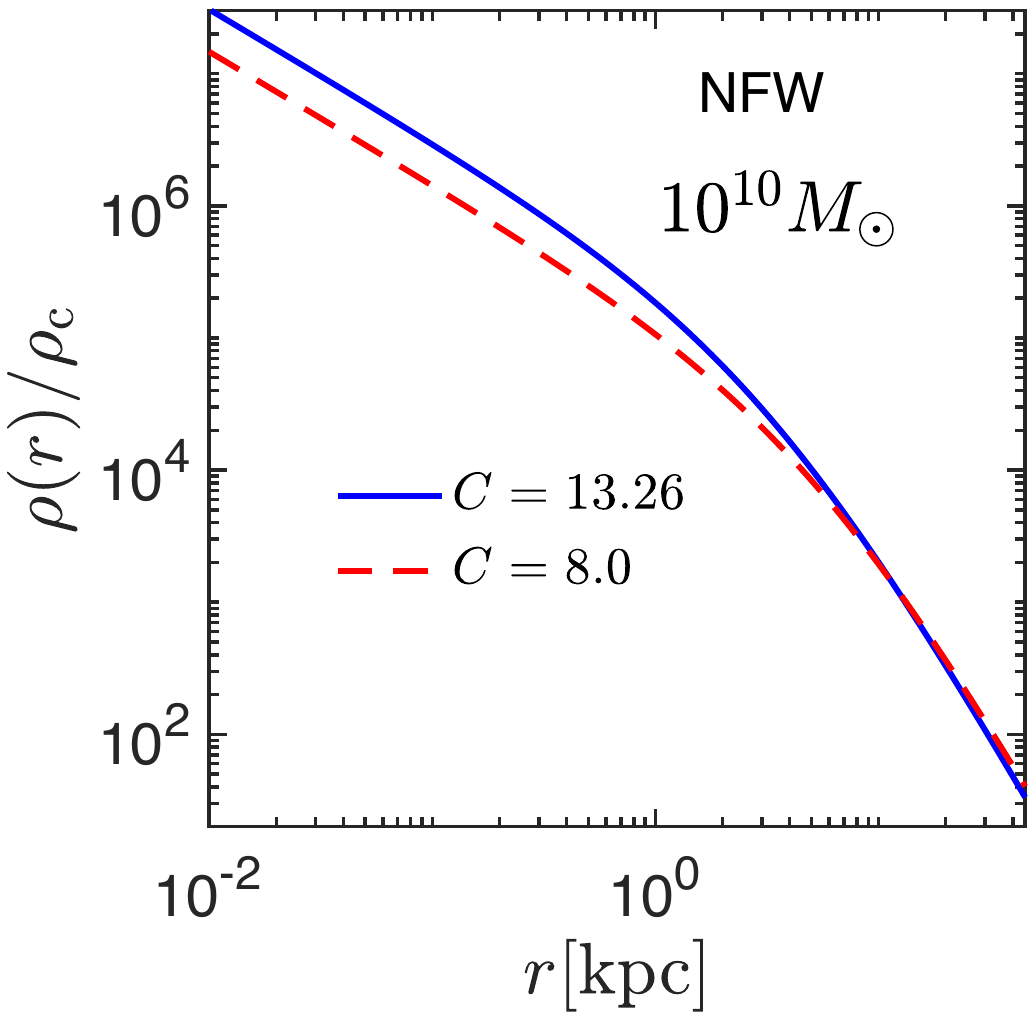}
  \vskip 0.1in
 \caption{ NFW mass density profiles for two  values of the concentration parameter for a halo of mass $10^{10}M_\odot$ ($r_{200}=46$ kpc). The high concentration corresponds to the mean  value obtained in N-body simulations, while the other  value is  40\% lower. }
\label{fig:mprof}
\end{figure}

 It is instructive to examine the  difference between the low and high concentration NFW profiles considered here.  Fig.~\ref{fig:mprof} plots  profiles  for a  halo with
 $M_{200}=10^{10}M_\odot$, for $C=\bar C\approx 13.26$ and a 40\% lower value, $C\approx 8$.
The virial radius is $r_{200}=46$ kpc,  but the profiles are close to each other down to $r\sim 5 $ kpc, which is 
roughly the radius of the inner region probed by the observed kinematics of the 10 star clusters in \NDF . 
\subsection{Tidal stripping/trimming: reduction of velocity dispersion}
 
Crucial to the mass modeling is the assumption that \NDF\   had been stripped of matter outside a radius of $\sim 10$ kpc, by the tides of external gravitational forces. As we will see below, the net 
effect of that is  a reduction in the velocity dispersion, while maintaining almost the same amount of interior mass. 
At   $D\approx 20$ Mpc,  \NDF\   is likely associated with  the much  larger galaxy  NGC 1052 at a projected distance of $\approx 75$ kpc. This large elliptical galaxy is a reasonable candidate for 
supplying the necessary gravitational tides.  
 It should be pointed out that the relative radial  velocity of \NDF\ relative to  NGC 1052 is 
$293\kms$ \citep{vanDokkum2018}, while  the l.o.s  velocity dispersion of the NGC 1052 group is only $110\kms$  \citep{vanDokkum2018}, which is consistent with the   circular velocity  of $\vc=200\kms$ measured from the \hi\ content of NGC 1052  \citep{vanGorkom+86}.  Thus the relative speed of 
\NDF\ is close to the escape velocity from NGC 1052.
At a projected distance of $100$ kpc between  the two galaxies, \NDF\   is likely to be just skimming past NGC 1052. However, a larger mass for  NGC 1052 is  obtained from   the SHMR relation. 
For    NGC 1052, the   stellar mass  of  $\sim 10^{11}M_\odot$ \citep{2017MNRAS.464.4611F}, translates 
to a 
 halo  mass of $\sim 5\times 10^{12}M_\odot $ \citep{Wasserman2018} according to the SHMR of  \cite{Rodriguez-Puebla2017}.
 In any case, there is a large uncertainty in the estimation of the tidal radius of \NDF\ in the presence of NGC 1052 \citep{Ogiya2018, Wasserman2018}.  Nevertheless,  a tidal radius of  $\sim10$ kpc is consistent with  the observed  spatial extent of the stellar component of \NDF\ and the distribution of the projected distances of its star clusters.

Particles with high apocenters reaching the inner regions move faster than those bound to inner orbits. Thus, excising the outer particles,  narrows the range of 
velocities in the inner regions. 
To illustrate this point, without offering a realistic quantification of the effect, we resort to a simulation of an isolated   halo of mass  $M_{200}=1.3\times 10^{10}M_\odot$. The simulation is similar to those used by 
\cite{Nusser2018} to study the effect of  dynamical friction  of the population of the star clusters in \NDF . The  simulation contains $2.4\times 10^5$  particles of equal mass, 
distributed  according to the Einasto mass profile. 
The particles  were advanced   forward for 10 Gyr using the publicly available \textit{treecode} by  \cite{Barnes1986}.
Further details can be found in \cite{Nusser2018}. 
We  select those particles found   at distances of less than a trimming radius $r_{\mathrm{tr}}=10$  kpc from the halo center in all 50  snapshots covering  the last 5 Gyr of the simulation. 
We assume that those particles were never at a farther distance and we identify them as composing the  halo surviving a tidal stripping process. 
We then project the particle velocities  along a single axis (the l.o.s) and 
compute the corresponding  velocity dispersion, $\sigma_u$,  at a 3D distance $r$ from the center. We compare this with the velocity dispersion  computed from all
 original particles (i.e. including pre-trimming particles  passing by from distances beyond  10 kpc). 
 
\begin{figure}
 \includegraphics[width=0.48\textwidth]{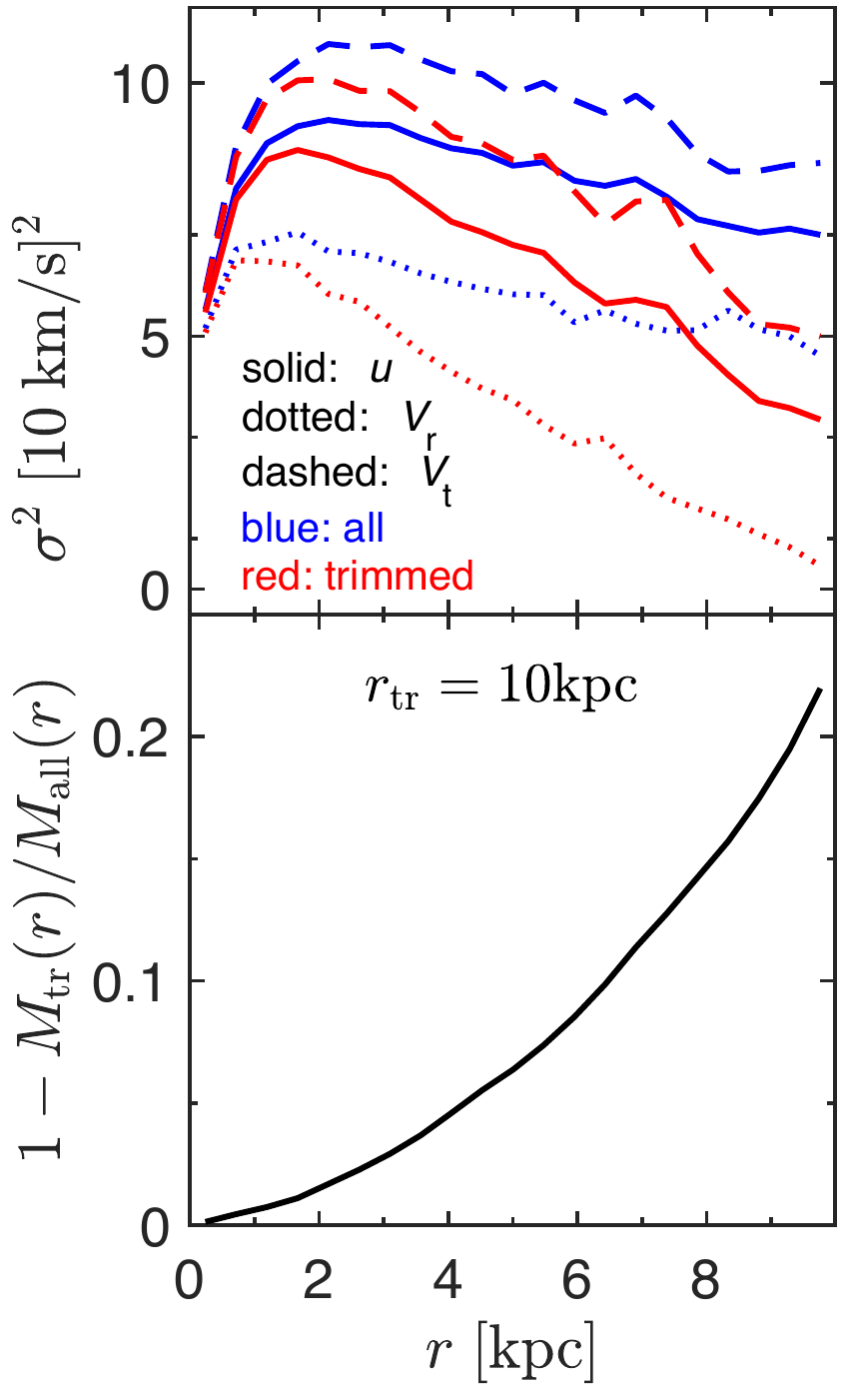} 
  \vskip 0.0in
 \caption{ \textit{Top:} The variance of velocity,  versus the 3D radius,  of particles inside a radius of $\rtr=10$ kpc  in a simulated halo of  $M_{200}=1.3\times 10^{10}M_\odot$. 
 The blue curve is the variance from  particles in the pre-trimmed halo, where the red curve is only for particles with apocenters smaller than $\rtr$.
 Solid, dotted and dashed curves correspond to velocities in l.o.s, radial and tangential directions, respectively. The variance of the tangential velocity is divided by $2$.
 \textit{Bottom: } Relative reduction of mass   within $r$ due to removal of  particles outside $10$ kpc.}
\label{fig:sigtrim}
\end{figure}

 The red and blue solid curves in the top panel of Fig.~\ref{fig:sigtrim} plot
 square velocity dispersions versus $r$  for the  trimmed (tidally stripped) and the untrimmed halos.  The l.o.s velocity 
 variance 
 $\sigma_u^2$ is represented by the solid curves. 
As $r$  approaches the trimming radius of $10$ kpc,  $\sigma_u^2$ of the trimmed halo (red solid  curve)  falls substantially below 
 the untrimmed halo (blue solid). 
 The ratio of $\sigma^2_u$   between the two curves 
approaches  unity   only at $r<2$ kpc.  The dotted curves represent the variance, $\sigma^2_r$, of the radial (from the halo center) velocity component. 
The red dotted curve  actually drops to zero at 10 kpc, since there are no particles moving beyond that radius. The dashed curves are the  tangential  velocity variance divided by 2,  $\sigma^2_t=(\sigma^2_\theta+\sigma^2_\phi)/2$.   Trimming also reduces the tangential dispersion but  the reduction  in the $\sigma^2_u$ is 
mainly due to the lowered dispersion in the radial direction. Note that the velocity dispersion ellipsoid in the original halo  is skewed to  the tangential direction,  with 
an anisotropy parameter  $\beta=1-\sigma_t^2/2\sigma_r^2 \approx -0.7 $ for $2 \ltsim r\ltsim 10$ kpc. The reduction in $\sigma_u$ clearly depends on the parameter $\beta$ of the original halo. 
We emphasize that the numerical experiment presented here is simply for illustration purposes. In fact, halos in simulations with  generic initial conditions based on the $\Lambda$CDM model, have a marked tendency 
at large radii towards a positive $\beta$ \citep[e.g.][]{2000ApJ...539..561C,Mamon2005,Ascasibar2008,Mamon2010,Lemze2012,Sparre2012}. This implies both a more significant  reduction in $\sigma_u$ and also a more isotropic velocity ellipsoid in the trimmed halo at $r<\rtr$.

\begin{figure}
 \includegraphics[width=0.48\textwidth]{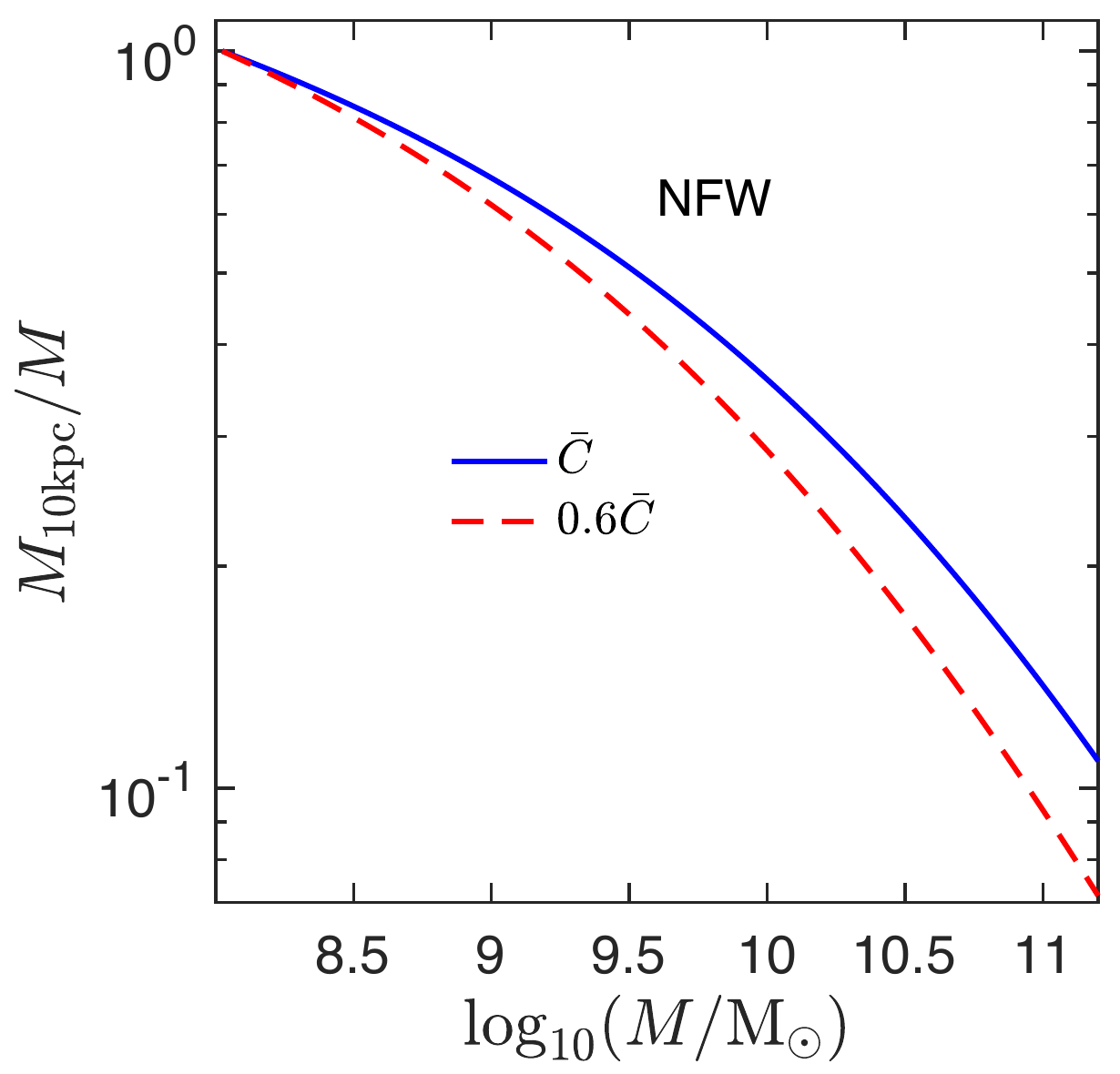} 
  \vskip 0.0in
 \caption{ The ratio of the halo mass, $M_\mathrm{10kpc}$, contained inside  a 10 kpc (3D) distance,  to the mass $M=M_{200}$ of the halo. The solid and dashed curves correspond to  $C=\bar C(M)$ and $0.6 \bar C(M)$, respectively. For halos with $M<10^8M_\odot$ ($r_{200}<10$ kpc) we set $M_\mathrm{10kpc}=M$. }
\label{fig:M10}
\end{figure}

The trimming procedure has a much lesser effect on the mass, $M(r)$,   encompassed within a radius $r<\rtr$.
The bottom panel of Fig.~\ref{fig:sigtrim},  plots the reduction  in the relative mass between the trimmed and the pre-trimmed  halos.   The reduction is meager and does not exceed 22\%,  even at $r=10$ kpc. 
 Thus, in  our modeling of  $F_u$ we will ignore this mild reduction in the mass inside $\rtr$ and treat the gravitational force field according to the original 
 NFW mass profiles. The effect of tidal stripping  will simply be included by imposing a  vanishing density outside $\rtr$. 
 Tidal stripping is a cumulative process which can be modeled accurately  for a known host mass profile and  the orbit of the satellite galaxy. Unfortunately, none of these are  accessible to us 
 from current observations and therefore, in our ``semi-analytic" approach here, we simply perform a sharp  trimming of the halo at a certain  radius. 
 Since our modeling is simplified we do not term this radius the tidal radius as is common but rather the trimming radius. 
 In any case,  gravitational tidal stripping always results in a radius beyond which the density effectively drops to zero. 
Fig.~\ref{fig:M10} gives the useful information of the fraction of the DM halo mass, $M_\mathrm{10kpc}$  within $10$ kpc relative to the virial halo mass. 

\subsection{Distribution of tracers}
In addition to the mass model, the calculation of the  likelihood function, $\cL$,  in Eq.~\ref{eq:like} involves 
an assumed form for the distribution of tracers, $ \nu(r)$.
In \cite{vanDokkum2018} and \cite{Trujillo2018}, the distribution of the 10  tracer star clusters  in \NDF\  is represented as a power law where the cumulative distribution up-to a distance $r$ is 
$n(r)\propto r^{3-\gamma}$ corresponding to $\nu \propto r^{-\gamma}$.
We follow these authors and adopt this form.


\section{Results}
 \label{sec:results}

This section begins with  a general illustration of the shape of predicted DFs. Then it describes the way  we  ascertain the validity of our analysis framework and 
how  confidence levels (CLs)  on the relevant parameter are derived. Then, we  present the actual implications of the data on the halo mass.

\subsection{Predictions for  the  DF}

\begin{figure*}
  \vskip 0.2in
 \includegraphics[width=0.8\textwidth]{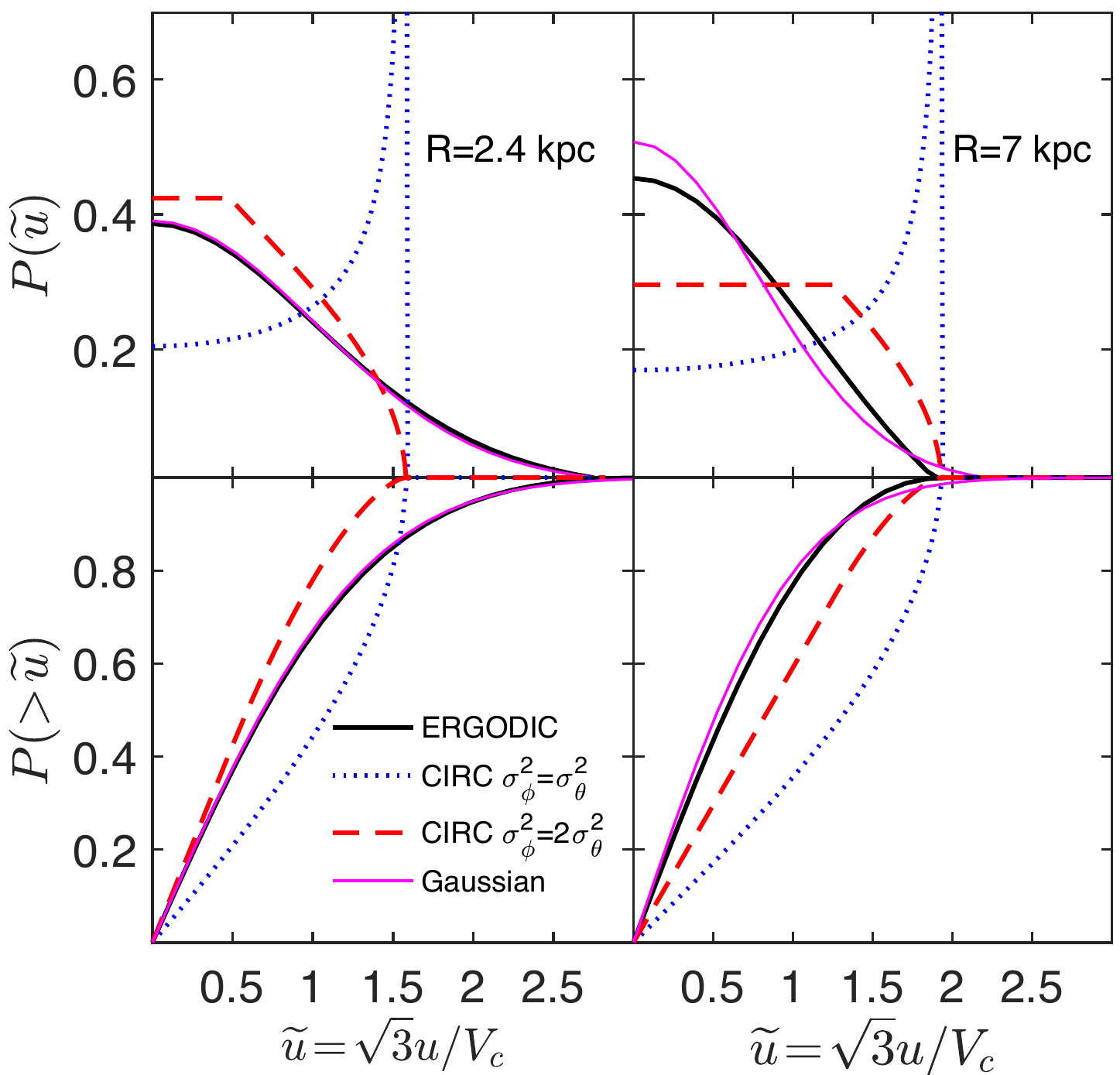}
  \vskip 0.1in
 \caption{ The PDF of the  l.o.s velocity $u$ in units of $V_c/\sqrt{3}$ for tracers at projected distances $R=2.4$ kpc (left) and 7 kpc (right). Three models are plotted: ergodic (black solid), circular orbits with isotropic angular distribution (dotted blue) and
 a non-isotropic distribution (dashed red). For comparison, a Gaussian PDF is also shown (solid magenta). The figure corresponds to 
a  DM halo with $ M_{200}=1.3 \times 10^{10}  $, $V_c=V_{200}$, and  $\gamma=2.4$.  \textit{See text}.}
\label{fig:Fu}
\end{figure*}

The mass model dictates $F(u,R) $ which, in turn determines  the likelihood 
$\cL$ through Eq.~\ref{eq:like}. It is thus instructive to examine  the   behavior of  $F(u,R) $ in some detail. Fig.~\ref{fig:Fu} shows  curves of $F$  in terms of $\widetilde u=\sqrt{3}u/V_c$, where $V_c=GM_{200}/r_{200}$.
The calculations are done for halo with  $M_{200}=1.3\times 10^{10}$ ($V_c=34\kms$), $C=0.6 \bar C=7.7$ and  a power law index of the  spatial distribution of tracers with   $\gamma=2.4$. This choice for $\gamma$ is motivated by the subsequent analysis and is consistent with \cite{Trujillo2018}. The  trimming radius is $\rtr=10$ kpc, encompassing  $\sim 4\times 10^9M_\odot$ in 
DM (c.f. Fig.~\ref{fig:M10}), in addition to the stellar content. The set of equations in \S\ref{sec:ergodic} is then used to calculate
the DF under the ergodic assumption. The DF for  circular orbits with isotropic angular distribution is  obtained using  Eq.~\ref{eq:circiso}. 
We also consider a non-isotropic angular distribution of orbits corresponding to  $\sigma^2_{\phi}=2\sigma^2_{\theta}$ and  Eq.~\ref{eq:circ}.
In all these three cases, Eq.~\ref{eq:fuF} is employed to obtain the DF in terms of the projected distance $R$ rather than the 3D distance $r$.
The calculation of the gravitational potential $\psi$ is done by numerically  integrating the mass profile on a radial grid and the derivative  $\dd \nu / \dd \psi$ is then obtained by 
means of numerical differences. 

 The figure shows 
  the  PDF  $P(\widetilde u)$ obtained from the DFs  at a projected distance  $R=2.4$ kpc and $7$ kpc, as indicated. The PDFs are symmetric with respect to the sign of $u$ and hence the figure refers only to positive velocities. The ergodic DF, shown as the black solid curve, declines with increasing velocities  and matches a Gaussian for $R=2.4$ kpc. However,  closer to $\rtr$, at $R=7$ kpc, there is a clear deviation from a Gaussian. At this distance, there is a cutoff at  $u\approx V_c$ resulting from the maximum velocity a tracer can have without crossing $\rtr=10$ kpc.  The circular orbit  PDF with isotropic distribution (dotted blue) has a distinctive    dip at $u\sim 0$, diverges   
  around the circular velocity of the halo, $V_c$,  and vanishes at larger velocities. The PDF for anisotropic circular orbits (red dashed) is constant at low velocities, in accordance with  Eq.~\ref{eq:circ}. However, because of the integration over the l.o.s (see Eq.~\ref{eq:fuF}), the PDF declines 
  at large velocities.  We already can see that the mass constraints will be lowest for the assumption the isotropic circular orbits.  For circular orbits, 
  the trimming radius affects $F$ only through the integration over $z$ in Eq.~\ref{eq:fuF}. 
  
\begin{figure*}
\vskip .2in
 \includegraphics[width=0.96\textwidth]{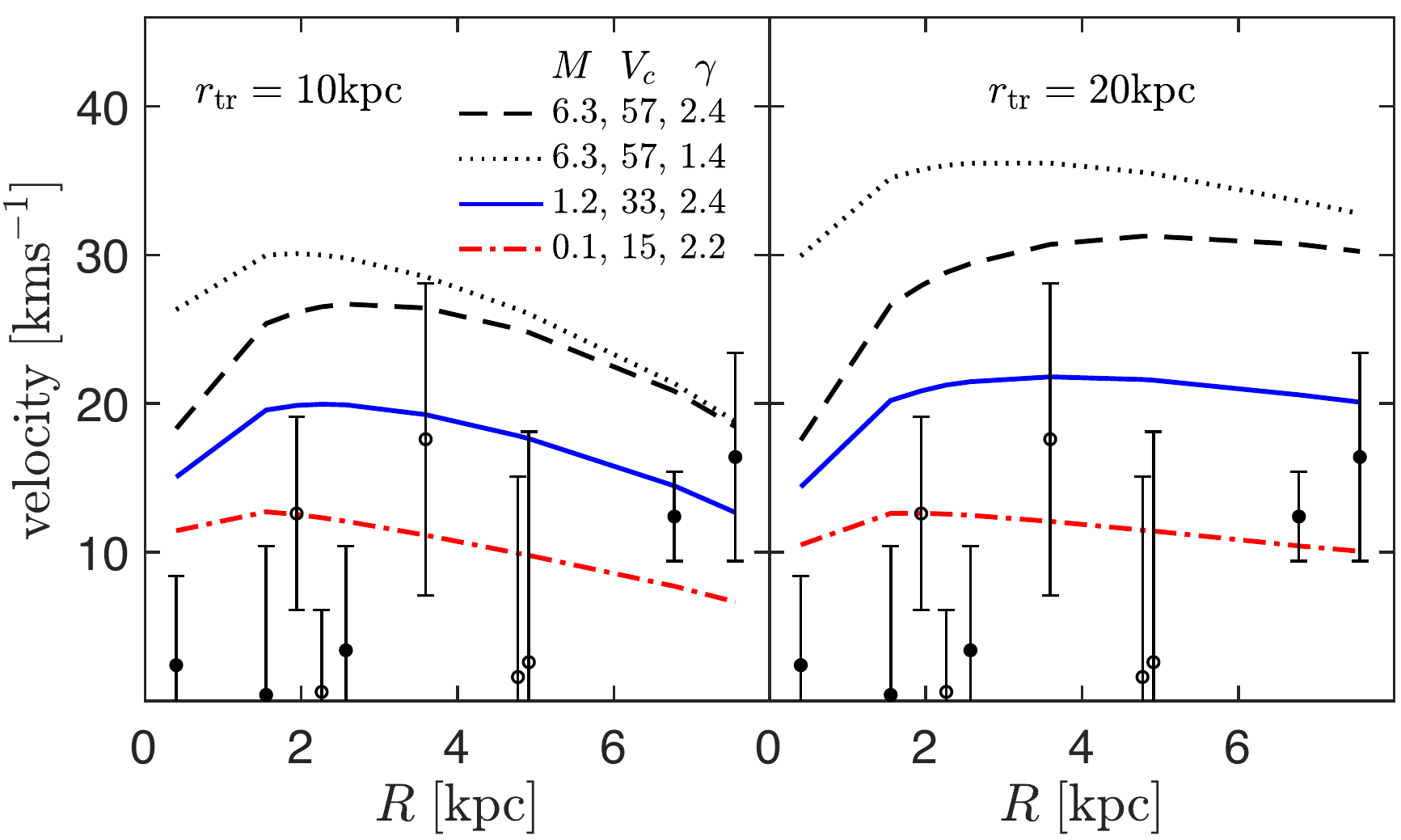}
  \vskip 0.1in
 \caption{The predicted l.o.s velocity dispersion from the ergodic DF, versus  projected distance, an NFW profile with $C=0.6\bar C(M)$. 
 The \textit{left} panel  for a trimming radius $r_\mathrm{trim}=10$ kpc,
while   $r_\mathrm{trim}=20$ kpc  is plotted in the  \textit{right} panel. Presented are results for 3 choices of  the  halo mass $M=M_{200}$, as indicated in the figure  in units of $10^{10}M_\odot$. 
The corresponding halo circular velocity $V_c$ is in $\kms$. For the highest mass, we plot curves representing two values of $\gamma$, 
as indicated.
Points with error-bars are the observed velocities and $1\sigma$ error-bars from 
Wasserman et al. (2018), where filled
and open circles correspond, respectively, to positive and  negative velocities relative to the mean. 
 The error bars do not include uncertainties due to the systemic velocity. }
\label{fig:udata}
\end{figure*}

   Fig.~\ref{fig:udata} plots the measured \NDF\ l.o.s. velocities and the predicted   velocity dispersion, $\sigma_u^2$, from the ergodic DF for an NFW mass profile with 3 different masses, as indicated in the figure. The prediction depends on $\gamma$ and hence we plot results for two values of $\gamma$ for the most massive halo considered in the figure.
 As expected, the large trimming radius, $\rtr=20$ kpc, (right panel) is associated with a higher  $\sigma_u^2$ than $\rtr=10$ kpc (left panel).  Comparison between the two panels, 
 reveals that  trimming has a substantially more pronounced effect on  the more massive halos. For the halo with $M_{200}=6.3\times 10^{10}M_\odot$, the dispersion at $R=6$ kpc  drops from $30\kms$ (left panel) to $22 \kms$ (right) for $\gamma=2.4$ (see dashed curves).  For  smaller halos, the trimming radius is closer to the virial radius resulting in a much smaller effect. 
 The reduction in $\sigma_u$ is almost unnoticeable for $M=10^9M_\odot$ (red dash-dotted), 
 where, according to Fig.~\ref{fig:M10}, $M_\mathrm{10kpc}$  is as high as $80\%$ as the original mass, $M$. The dependence on $\gamma$ is actually significant, as seen by comparing the dashed and dotted curves in each panel. 

\subsection{Confidence levels and the goodness of fit}
The CLs on  model parameters will be set
 according to the likelihood ratio in terms of  $\Delta \tilde \chi^2=-2\ln(\cL/\cL_\mathrm{m})$, where $\cL_\mathrm{m}$ is the maximum $\cL$ over the parameters considered.

Before presenting the results for the CLs,  we would like to  confirm that   parameters  corresponding to $\cL_\mathrm{m}$ are indeed a good fit to the data-- otherwise the whole analysis framework is questionable. An indication for the validity of our analysis can already be 
seen in Fig.~\ref{fig:udata} where the dot-dashed curve appear to agree  well the velocity measurement. In the following we aim at a more robust validation of our  framework.

\begin{figure}
  \vskip 0.2in
 \includegraphics[width=0.49\textwidth]{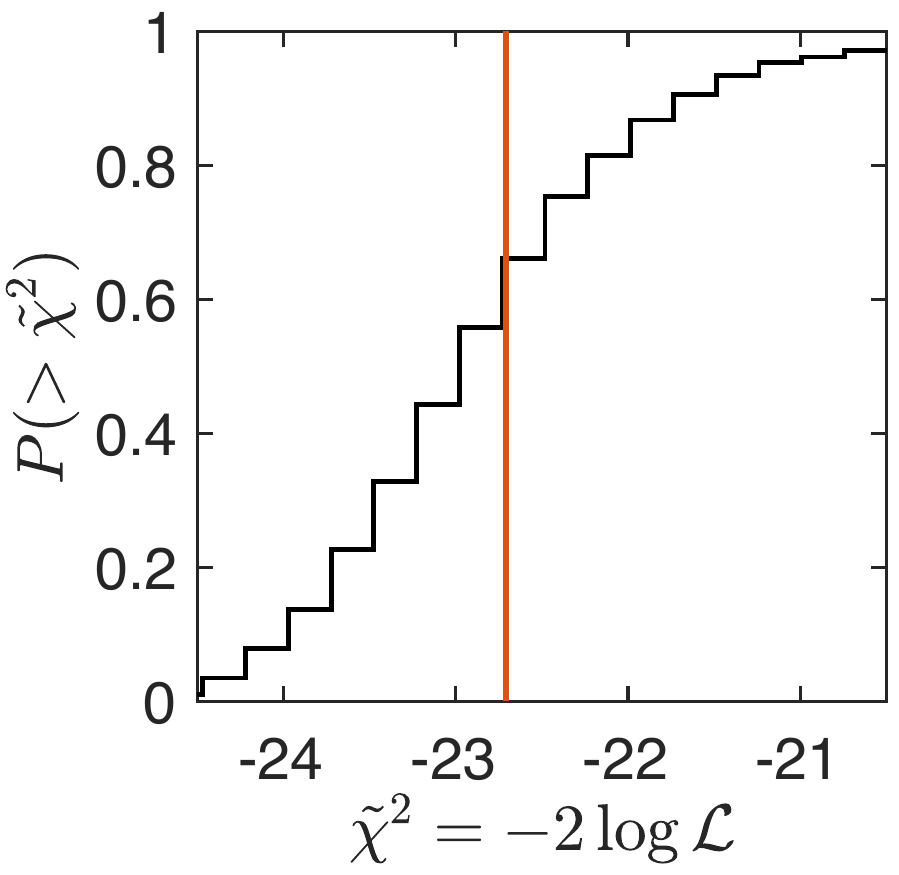}
  \vskip 0.1in
 \caption{The cumulative distribution of ${\tilde \chi}^2$ obtained from $10^3$ random mock data sets drawn from the ergodic $F$
 for an NFW halo with $M_{200}=5\times 10^8M_\odot$ with a concentration $C=0.6\bar C$. The trimming radius is $10$ kpc. The vertical line indicate the value obtained from the observations for the same halo model. }
\label{fig:chin}
\end{figure}

As it will be seen  below (see Fig.~\ref{fig:chiERG}),  a minimum of $\Delta \tilde \chi^2$  is obtained at $\gamma_\mathrm{min}\approx 2.3$ and $M_\mathrm{min}\approx  5\times 10^{8}M_\odot$ for an NFW mass model with $C=10.7$ 
(40\% lower than the mean concentration for that mass). 
Using these values,  we generate such mocks by  
drawing random $\vz$  from the DF $F(\vz,R)$.
 Random velocity errors are then added to $\vz$ to obtain $\vz_0$. 
This way, we produced a large number of mock catalogs each containing the same number of tracers as the observations. The mock catalogs are then treated as the real data, yielding a $\cL$ value for each mock.  As a measure of the goodness-of-fit, we examine 
the agreement  of the observed  $\cL_\mathrm{m}$ with the distribution of $\cL$ from the mocks. 
Fig.~\ref{fig:chin} gives the cumulative distribution of $\tilde \chi^2=-2\ln\cL $ from $10^3$ mocks corresponding to the 
ergodic DF. The vertical line in the figure, indicating
$\cL_\mathrm{m}$  from the observations,  is clearly  consistent with the distribution from   mocks, with a $\ltsim 1\sigma$ from the mean. 
The result  strongly validates  our basic modeling framework.
\subsection{The likelihood $\cL$ as a function of  $M$  and $C$}
 \label{sec:MC}
\begin{figure}
  \vskip 0.2in
 \includegraphics[width=0.48\textwidth]{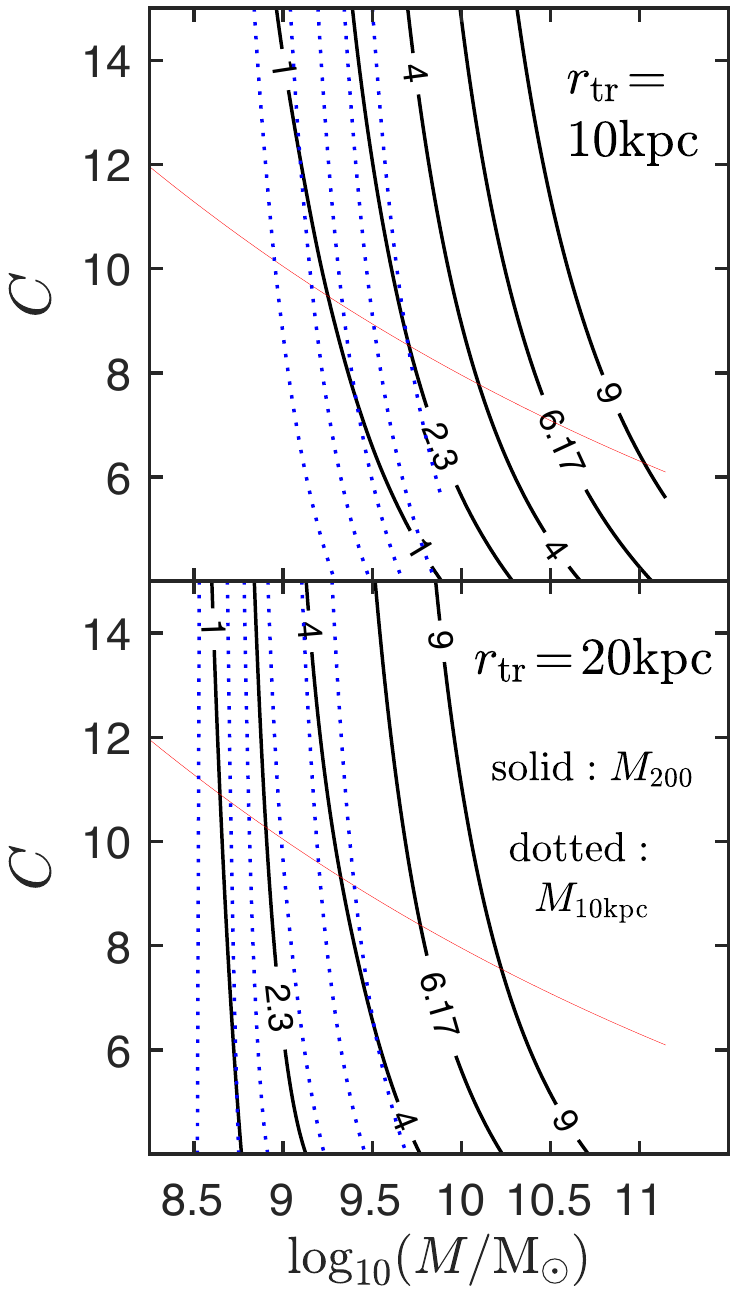}
  \vskip 0.1in
 \caption{Contour maps of $\Delta {\tilde \chi}^2$  in the plane of  the concentration $C$ and mass. 
In both panels, solid and dotted contours correspond, respectively, to halo mass $M_{200}$ and $M_\mathrm{10kpc}$ (halo + stellar mass). The top and bottom panels represent  results for  NFW profiles trimmed, respectively, at $\rtr=10 $ kpc and $20$ kpc. 
 The results are obtained assuming an ergodic DF and a  $\gamma=2.4$ and the thin red curve represents $0.6\bar C(M)$.
  Numeric values are indicated on the solid contours only. Dotted contours correspond to the same CLs.}
\label{fig:contC}
\end{figure}

 Assuming the ergodic DF, we calculate   $\Delta \tilde \chi^2$ as a function of the concentration and mass, with 
  fixed $\gamma=  2.4$ and  $\rtr=10$ kpc. 
The results are shown in Fig.~\ref{fig:contC}  by means of contour maps.  Only contours of  $\Delta \tilde \chi^2=1$,  2.3,  4,  6.17 and  9 are plotted. For  two degrees of freedom, as our case is, 
 these contours correspond, respectively,  to CLs of  39\%,
68\% ($1\sigma$ for 1D normal distribution), 
86\% ($1.52 \sigma$) 
95\% ($2 \sigma$) 
and 98.9\% ($2.54 \sigma$).
 The data probes  a limited region, $R\ltsim 8 $ kpc, associated with a small fraction of the halo mass (see Fig.~\ref{fig:M10}).
Unsurprisingly, this leads to a degeneracy between the $C$ and either $M_\mathrm{10kpc}$ or 
the virial mass. The degeneracy is naturally more pronounced for $M_\mathrm{10kpc}$, as 
implied by the steeper vertical orientation of the blue dotted contours.

The degeneracy can be broken by multiplying $\cL$ by a  prior PDF
of  $C$  \citep{Wasserman2018},   taken from the concentrations measured from halos in simulations. However, constraints on $C$ obtained this way, are  entirely driven  by the prior. 
For the remainder of the paper, we  proceed in a  different  path, as described next.

\subsection{$\cL$ as a function of $M$ and $\gamma$}
We are primarily  motivated by  the question of whether the kinematics of  \NDF\    rules out  a parent halo with a mass  close to 
the  SHMR. 
According to Fig.~\ref{fig:contC},    smaller masses are associated with higher concentrations. Therefore, 
to assess the compatibility  a mass $M$   with the \NDF\ observations as well as the SHMR, it is sufficient to consider only an acceptable 
 low $C$ value.  
As suggested in \S\ref{sec:massmodel}, we set the low 
 value to  $C=0.6 \bar C(M) $ which falls below the mean relation, $C=\bar C(M)$ at the $\sim 1\sigma $, as seen for simulated halos. For comparison, we will also consider  $C=\bar C(M)$. 
The  power index, $\gamma$, of the assumed power law distribution of tracers affects the DF and 
thus we treat it as a second parameter, in addition to the mass.   
We proceed with  $\cL$ computed for  an array of $M$ and $\gamma$.
The lowest $M$ we consider is $10^8M_\odot$, close to the stellar mass of \NDF .

\begin{figure}
  \vskip 0.2in
 \includegraphics[width=0.48\textwidth]{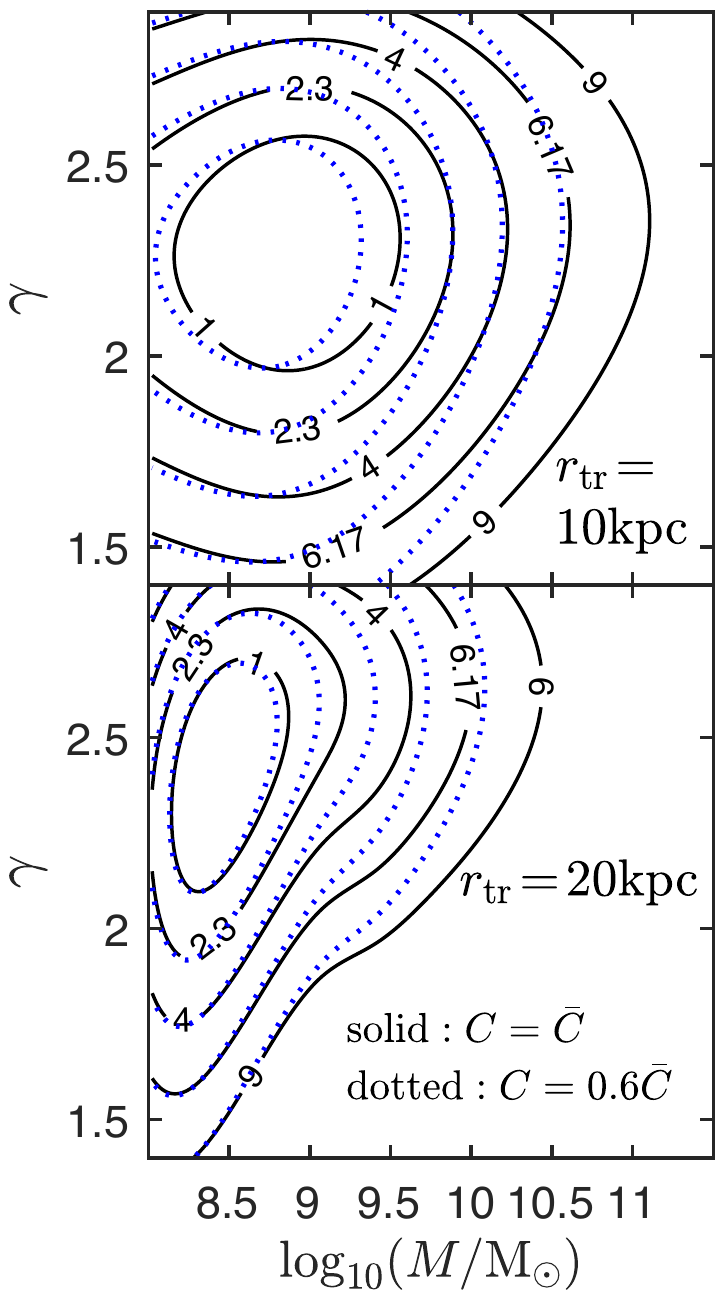} 
  \vskip 0.1in
 \caption{ Contour maps of $\Delta {\tilde \chi}^2$  as a function of   $\gamma$ and  $M=M_{200}$ obtained from the ergodic DF.
 Solid and dotted contours correspond, respectively, to $C=0.6\bar C(M)$ and 
 $C=\bar C(M)$.}
\label{fig:chiERG}
\end{figure}

\begin{figure}
\vskip 0.2in
 \includegraphics[width=.48\textwidth]{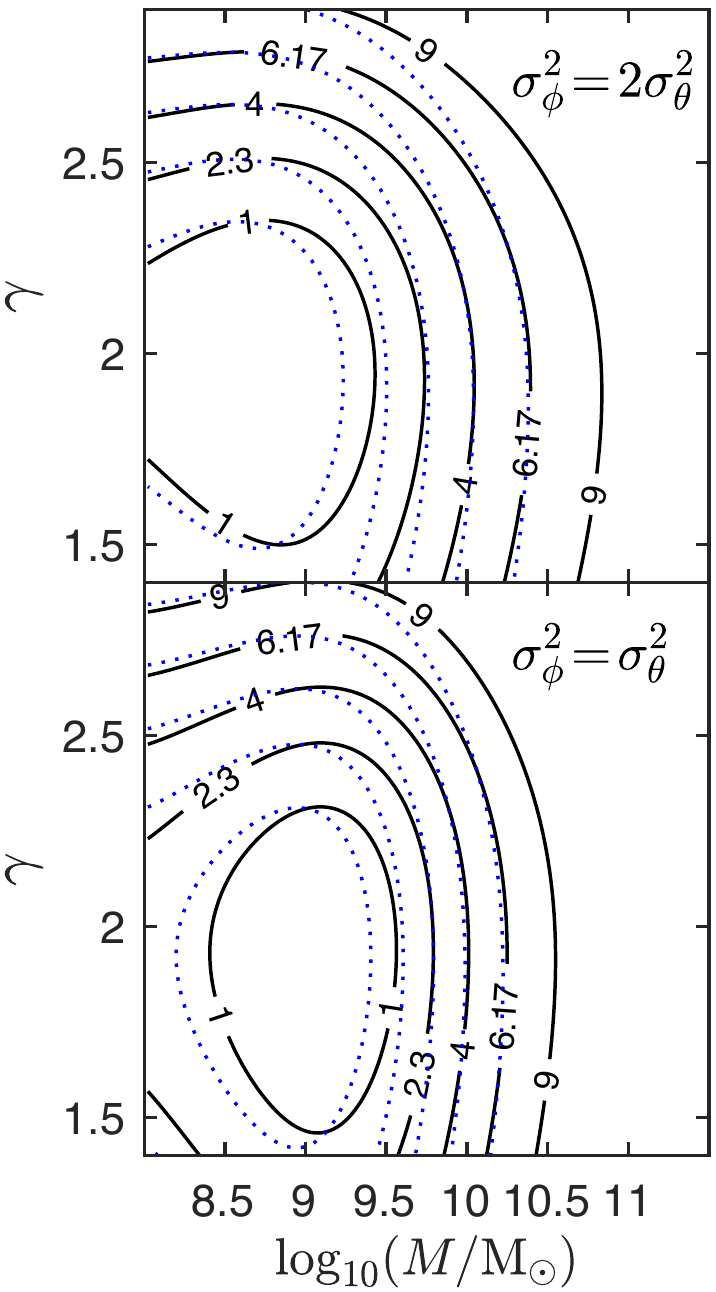}
\vskip 0.1in
 \caption{ Contour of $\Delta {\tilde \chi}^2$ for NFW mass profile obtained from the DF with circular orbits  for isotropic (bottom) 
and non-isotropic (top) angular distribution of tracers.  As in the previous figures, 
$\rtr=10$ kpc and, as before,  the solid and dotted contours  correspond to low and high concentrations . }
\label{fig:chicircNFW}
\end{figure}

  Contours of  CLs  in the    $\gamma-M$ plane are plotted in Fig.~\ref{fig:chiERG}. The figure refers to the ergodic DF, where the top 
  panel is for the  fiducial $\rtr=10$ kpc and the bottom is for  $\rtr=20$ kpc.
Solid and dotted contours in each panel represent NFW mass models with  $ C=0.6\bar C(M)$ 
 and $\bar C(M)$, as indicated in the figure.  As in Fig.~\ref{fig:contC}, CL contours for  5 values of $\Delta \tilde \chi^2$ are drawn. 
 As expected, the higher concentration value  (dotted curves) produces contours 
 which are more  squeezed to lower masses. Note that the $\Delta \tilde \chi^2=6.17$   solid contour ($C=0.6\bar C(M)$) roughly overlaps with $\Delta \tilde \chi^2=9 $    
 dotted curve ($C=\bar C(M$).  
 The bending up of  the contours with increasing $M$, 
indicates that  the constraints on $\gamma$ are  not merely  from the observed distribution of projected distances, but also through the kinematical effect of $\gamma$ on the DF.
Still, the dependence on $M$ is weak and the distribution of the projected distances is nonetheless  the main probe of $\gamma$. At $M=10^9M_\odot$ we find $\gamma\approx 2.33\pm 0.35 $, consistent with the result of \cite{Trujillo2018}. 
The trimming radius plays a pivotal role, with the larger $\rtr$ yielding a factor of $\sim 4$ lower masses for $\Delta \tilde \chi^2 =4$.
The break in the shape of the contours in the lower panel at $\log_{10}\approx 9$ reflects the fact that trimming does not affect small halos with  $r_{200}<20$ kpc. 
We also have calculated $\cL$  assuming circular orbits. The results are shown in Fig.~ \ref{fig:chicircNFW} where, in both panels,  the fiducial  $\rtr=10$ kpc is used in computing the CLs. 
In a spherical system, the 3D velocity of a particle on a circular orbit is determined by the enclosed mass. Thus, the dependence of DF of the observables and hence also the likelihood 
on the trimming radius, is weaker and arises  only from the geometrical projection effects.  
The bottom panel corresponds to orbits with random orientations and hence, an isotropic tangential velocity dispersions, with $\sigma_\phi^2=\sigma_\theta^2$ where, as in \S\ref{sec:circ}, 
$\theta$ and $\phi$ are  defined assuming the l.o.s is in the $ z$ direction. 
  In  the top panel, a mild anisotropy, 
in the orientation of the orbits  with respect to the  l.o.s., is invoked  ( cf.  \S\ref{sec:circ} for details). 
The anisotropy leads   leads to $\sigma_\phi^2=2\sigma_\theta^2$,    and subsequently a lowering of $\sigma_u^2$, in accordance with the curves of the DFs in Fig.~\ref{fig:Fu}. 

\subsection{Upper limits on the mass }
For the purpose of deriving upper limits on the mass, we consider  $\Delta \tilde \chi^2(\gm,M)$,  where  $\gm$ is the ``best-fit" value of $\gamma$ rendering a maximum 
 $\cL(\gamma, M) $ for a given  $M$. Thus we treat $\gamma$ as a ``nuisance" parameter using the  \textit{profile-likelihood }  approach\footnote{This approach is equivalent to \textit{ marginalization}  for normal forms of $\cL$ and 
 of   the assumed  prior PDF  of  $\gamma$.}.
 
\begin{figure}
\vskip .2in
 \includegraphics[width=0.48\textwidth]{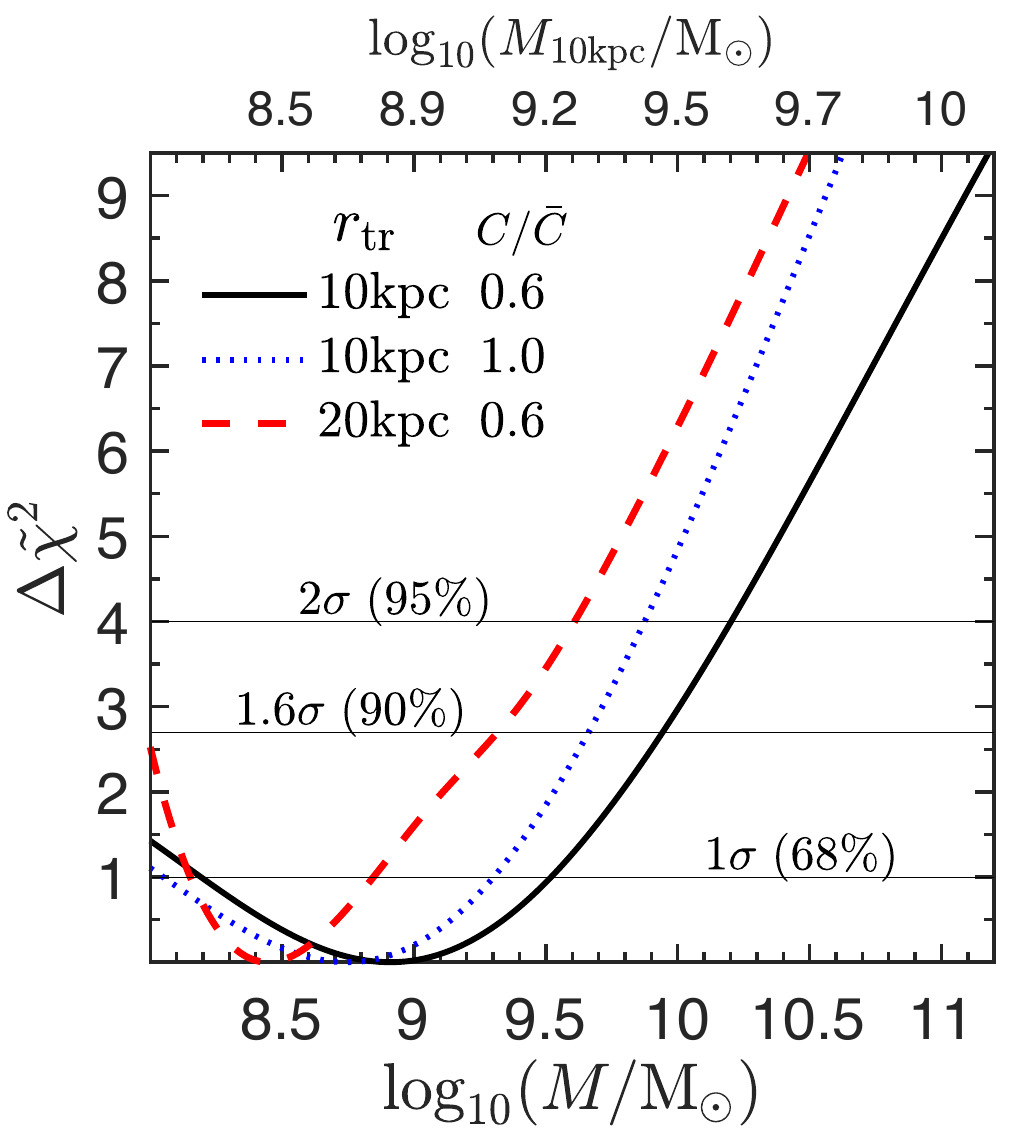}
 \vskip .1in
 \caption{ Curves of  $\Delta \tilde \chi^2 $ as function of mass, for the ergodic DF.
 Solid and dotted lines are for $\rtr=10$kpc but with  $C/\bar C(M)=0.6$   and 1, respectively. The red dashed curve is for $\rtr=20$ kpc and the low $C$ value.
 The bottom axis is the mass, $M_{200}$ while the top axis is the (DM+stellar) mass within 10 kpc for $C=0.6\bar C(M)$.
For a given mass, $\gamma$ is adjusted to its  best-fit value, $\gm$. Thin horizontal lines indicate CLs.}
\label{fig:chi1d}
\end{figure}

We seek upper limits on the DM halo mass for our  adopted  NFW profile with  $C=0.6\bar C(M)$.
 The solid curve in Fig.~\ref{fig:chi1d} plots  $\Delta \tilde \chi^2(\gm,M)$  for this profile. The dashed and dotted curves correspond, respectively, 
 to  results obtained with  $\rtr=20$ kpc  and $\rtr=10$ kpc with a high concentration. 
 Mass upper limits from these curves are given in the Table.

 For $\rtr=10 $ kpc, similar $M_\mathrm{10 kpc}$ are obtained for both choices of the concentration, although there is a factor of 
 2 difference in the inferred $M_{200}$. This is a reflection of the absence of mass tracers beyons $R\sim 10 $ kpc  and, therefore, the more observationally constrained quantity is $M_\mathrm{10 kpc}$.
 As we have seen in the contour maps, the results are highly sensitive to $\rtr$. Maintaining the low concentration, the  $1\sigma$ upper limit on the mass is lowered by a factor
 $\sim 20$ for  $\rtr=20 $ kpc relative to $\rtr=10$ kpc.

\begin{table}
\caption{Upper limits on the mass of \NDF\ . Masses are in units of $10^{9}M_\odot$.}
  \begin{tabular}{lcccccc}
    \toprule
              &  \multicolumn{6}{c}{($\rtr/\mathrm{kpc},C/\bar C$)} \\
    \multirow{2}{*}{CL} &
      \multicolumn{2}{c}{(10, 0.6)} &
      \multicolumn{2}{c}{(10, 1) } &
      \multicolumn{2}{c}{(20, 0.6)} \\
      & $M_{200}  $ & $M_\mathrm{10kpc}$ & $M_{200}  $ & $M_\mathrm{10kpc}$ & $M_{200} $ & $M_\mathrm{10kpc}$\\
      \midrule
    $1\sigma$  (68\%)     & 2.9 & 1.5     & 1.7 & 1.4      & 0.12 & 0.12 \\
    $1.64\sigma$  (90\%) & 8.1 & 2.7    & 4.3 & 3.3       & 1.8 & 1.1 \\
    $2\sigma$  (95\%)&    15.2 & 3.8       &  7.9 & 5.0    &   3.8 & 1.7 \\
    \bottomrule
  \end{tabular}
\end{table}
 
\section{Discussion and Conclusions}
\label{sec:discussion}

We have employed the phase space  distribution function to show that the kinematics of \NDF\  is compatible 
with a  dynamical  mass on the order of a few $10^{10} M_\odot $.  Our adopted  NFW mass model we adopt is consistent with 
the structure of halos found in cosmological simulations of the $\Lambda$CDM model. 
The model relies on  (possibly cumulative) action of tidal stripping events for the removal of matter beyond a distance $10$ kpc
from the center of \NDF .  
Our findings here are consistent with a model in which \NDF\ started as a  normal DM halo, but had 
its  matter beyond $\sim 10 $ kpc removed by external gravitational tides. 
 Alternatively, the progenitor of  \NDF\ could have always been under the influence of 
an external gravitational field, preventing its growth  beyond a certain radius. 
In both of these scenarios,   the  inner regions maintain most of their mass and do not undergo major 
disruption of  their  internal cold kinematics \citep{Ogiya2018,Wasserman2018}.

In a previous paper \citep{Nusser2018}, it was demonstrated  that a low dynamical mass for \NDF\ leads to short orbital decay times by dynamical friction. The high mass limits obtained here  resolve this issue by increasing the dynamical friction time scales by a factor 5 or so. 
However, the question arises whether the analysis, based on the assumption of a steady state, remains valid at low masses in light of the importance of dynamical friction. The dynamical time scale  for small masses is short compared to the Hubble time, but it is still longer than the dynamical time of the system. Thus we expect the analysis to remain valid even at small masses. We mention in passing that the  numerical tests of the previous paper, confirm that  dynamical friction has very little effect on the velocity dispersion. 
Recently,  \cite{Yu2018} identified  ``galaxies" in the  Illustris simulations \citep{Gene2014} having very little dark matter.
This is an  interesting finding,  but it may not actually be relevant to \NDF\ because of the short  dynamical friction time scale obtained for 
low dynamical mass.

\cite{2018MNRAS.tmp.2765L} argue that the estimated velocity dispersion from the 10 GC is biased low by about 10\%. Once this bias is incorporated, 
they find  an upper limit to the halo mass of $10^{10} M_\odot$ (95\% confidence). Despite the consistency of this mass estimate  with the our finding, 
their analysis is at odds with ours. In fact, with a large truncation radius,  the mass estimate derived from the full distribution done here yields results roughly consistent with  \cite{Wasserman2018} (see their figure 2). This indicates that the lower mass limits obtained by \citep{vanDokkum2018}  are not necessarily the result of  a bias in the estimation of the velocity dispersion \citep[see also][]{Martin2018}.

 \cite{Wasserman2018} and \cite{Hayashi2018} both  perfomed a Jeans analysis based on the velocity dispersion predicted 
from a mass model. They both assumed  a tracers'  distribution  
consistent with  a projected S\'ercic profile.  However, the conclusions of these two groups are different. \cite{Wasserman2018}  
find a dynamical mass-to-light ratio  of $1.7^{+0.7}_{-0.5} M_\odot/L_{\odot, V}$, while 
 \cite{Hayashi2018}  argue  that  the \NDF\ mass remains uncertain due to the unknown distribution of tracers. 
 \cite{Hayashi2018}   also find that with  S\'ersic based tracers' distribution, a boost by  a factor of 5  in  the mass within $7.6 $ kpc is achieved   in comparison  to \citep{Wasserman2018}.
Here, we opted to work with  a power law for  form $\nu\sim r^{-\gamma}$ and estimate $\gamma$ using the likelihood function. 
Given the small number of star clusters used as mass tracers, we feel that this simple form is quite  adequate. A much more 
important effect is related to the reduction of velocity dispersion by the removal of fast particles penetrating the inner regions from 
the outer regions. 

Recently, \cite{Emsellem2018} have derived a velocity dispersion of $16\pm 5 \kms$ for  the stellar component of \NDF\ using the  MUSE@VLT 
spectrograph. Although  the wavelenghth-dependent spectral resolution of the MUSE datacube is $\ltsim 35 \kms$, these authors could still derive a stellar velocity dispersion of $16\pm 5 \kms$. This certainly brings the virial mass comfortably close to the SHMR, but we should bear in mind that the spectral resolution is substantially larger than the inferred dispersion. 

Our  $ 2\sigma$ upper limit of $1.5\times 10^{10}M_\odot$  on the  mass of \NDF\ is  a factor $4-5 $ below the  SHMR \citep{Behroozi2010,Moster2013, Rodriguez-Puebla2017},  where at  the relevant mass scale, the scatter in the relation is about 0.36 dex \citep{Wasserman2018}. 
We should also keep in mind that the inference of  halo masses under the assumption of sphericity and steady state is associated with 
a factor of 2-3  uncertainties \citep[e.g.][]{Wang2017}. 
Thus we conclude that the possibility that \NDF\  has a high mass consistent with the SHMR cannot be ruled out.

\section*{Acknowledgements}
The author is grateful to Asher Wasserman for providing him with the l.o.s velocities and projected distance of the 10 star clusters.
Special thanks are due to the referee, Gary Mamon, for many comments which helped improve the manuscript. 
This research was supported by the I-CORE Program of the Planning and Budgeting Committee,
THE ISRAEL SCIENCE FOUNDATION (grants No. 1829/12 and No. 203/09)
and the Asher Space Research Institute. The author  acknowledges the hospitality
 of Department of Astronomy- Shanghai Jiao Tong University. 

\bibliographystyle{mnras}
 \bibliography{FVZ.bbl}
\end{document}